\documentclass[aps,prl,twocolumn,showpacs,superscriptaddress]{revtex4-2}
\usepackage{graphicx}
\usepackage{amsmath}
\usepackage{multirow}
\usepackage{tabularx}
\usepackage{color}
\usepackage{xcolor}
\usepackage{ulem}
\usepackage{float}
\usepackage{appendix}
\usepackage[colorlinks,linkcolor=blue,urlcolor=blue,anchorcolor=blue,citecolor=blue]{hyperref}

\newcommand{\dx}{{d_{x^2-y^2}}}
\newcommand{\dz}{{d_{3z^2-r^2}}}
\newcommand{\px}{{p_x }}
\newcommand{\py}{{p_y }}
\newcommand{\pz}{{p_z }}

\begin{document}

\title{ Charge Transfer and Zhang-Rice singlet bands in the Nickelate Superconductor $\mathrm{La_3Ni_2O_7}$  under Pressure}

\date{\today}

\author{W\'ei W\'u}
\email{wuwei69@mail.sysu.edu.cn}

\author{Zhihui Luo}

\author{Dao-Xin Yao}

\author{Meng Wang}

\affiliation{Center for Neutron Science and Technology, Guangdong Provincial Key Laboratory of Magnetoelectric Physics and Devices,
 School of Physics, Sun Yat-sen University, Guangzhou, Guangdong 510275, China
}

\begin{abstract}Recently, a bulk nickelate  superconductor $\mathrm{La_3Ni_2O_7}$ is discovered at pressures with a remarkable high transition temperature $T_c \sim 80K$. Here, we study a Hubbard model with tight-binding parameters derived from \textit{ab initio} calculations of $\mathrm{La_3Ni_2O_7}$, by employing  large scale determinant quantum Monte Carlo and cellular dynamical mean-field theory.
Our result suggests that the  superexchange couplings in this system are comparable to that of cuprates.
The system is a charge transfer insulator as hole concentration becomes four per site at large Hubbard $U$.  Upon hole doping,  two low-energy spin-singlet bands  emerge in the system exhibiting distinct correlation properties: while  the one composed  of the out-of-plane Ni-$d_{3z^2-r^2}$ and O-$\pz$ orbitals demonstrates strong antiferromagnetic correlations and narrow effective bandwidth,  the  in-plane singlet band consisting of the Ni-$d_{x^2-y^2}$ and O-$p_x / \py$ orbitals is in general more itinerant.  Over a broad range of hole doping, the doped holes occupy primarily  the  $d_{x^2-y^2}$ and $\px / \py$ orbitals, whereas the $d_{3z^2-r^2}$ and $p_z$ orbitals retain underdoped.  We propose an effective $ t-J$ model to  capture the relevant physics and  discuss the implications of our result for  comprehending the $\mathrm{La_3Ni_2O_7}$ superconductivity. 
\end{abstract}

\maketitle

\section{Introduction}\label{sec:intro}
 
Since the discovery of  cuprate superconductors~\cite{bednorz1986},  understanding and searching for novel high transition temperature (high-$T_c$) superconductors~\cite{sun2023, luo2023bilayer,  zhang1988, kotliar88, emery1995, scalapino2007, keimer2015quantum, maeno1994superconductivity,kamihara2006iron,maeno2001intriguing,chen2008superconductivity,chubu2008,hu2015predicting, Kitatani23,jiang2021ground,christos2023} has been one of the major focuses in the condensed matter physics. The discovery of infinite layer nickelate superconductor ~\cite{li2019, pickett04,middey2016physics,haule2017mott,peil2019,jiang2020critical,botana2020,zhang2020,werner2020,wang2020hund,lech2020,worm22,gu2020single,chen2022dynamical} marks a notable recent advancement, although in which the superconductivity (SC) has  been only observed in thin films on substrates~\cite{li2019}, but not yet in bulk samples~\cite{ding2023critical}. The very recently  discovered bulk  superconductor $\mathrm{La_3Ni_2O_7}$  under high pressures~\cite{sun2023} , which exhibits a remarkable $T_c$  of $\sim$ 80 Kelvins, then represents a significant breakthrough in this field.
As revealed by the density-functional theory (DFT) calculations~\cite{luo2023bilayer, sun2023,pardo2011},  a hallmark of the nickelate bi-layer  $\mathrm{La_3Ni_2O_7}$ ~\cite{sun2023, liu2023evidence} is the activating of  both 3$\dx$ and 3$\dz$ orbitals in vicinity of Fermi level~\cite{luo2023bilayer,gao22015}. This distinct feature may lead to superconductivity that differs significantly from the cuprates and infinite layer nickelates. From a theoretical perspective, several crucial questions arise. First,  to  understand the driving force behind the SC, it is necessary to elucidate the magnetic exchange couplings~\cite{jiang2020critical,lu2021magnetic} among the four active $e_g$ orbitals in the $\mathrm{NiO_2}$ bi-layer. Moreover, the $e_g$ orbitals of  $\mathrm{La_3Ni_2O_7}$ under pressure possess  a  large hole content,  as
the nominal valence of $\mathrm{Ni}$ is $\mathrm{Ni}^{2.5+}$ in this system, indicating  an average of $1.25$ holes per $e_g$ orbital.  This high  hole concentration level  is on the verge of quenching SC by overdoping in the context of cuprates. Therefore, resolving the distributions of the holes in the Ni-$e_g$ orbitals and  the correlated O-2$p$ orbitals is crucial for  understanding the correlation effects in $\mathrm{La_3Ni_2O_7}$ under pressure~\cite{Liu2023wen}.

\begin{figure}[htp]
\includegraphics[scale=0.35]{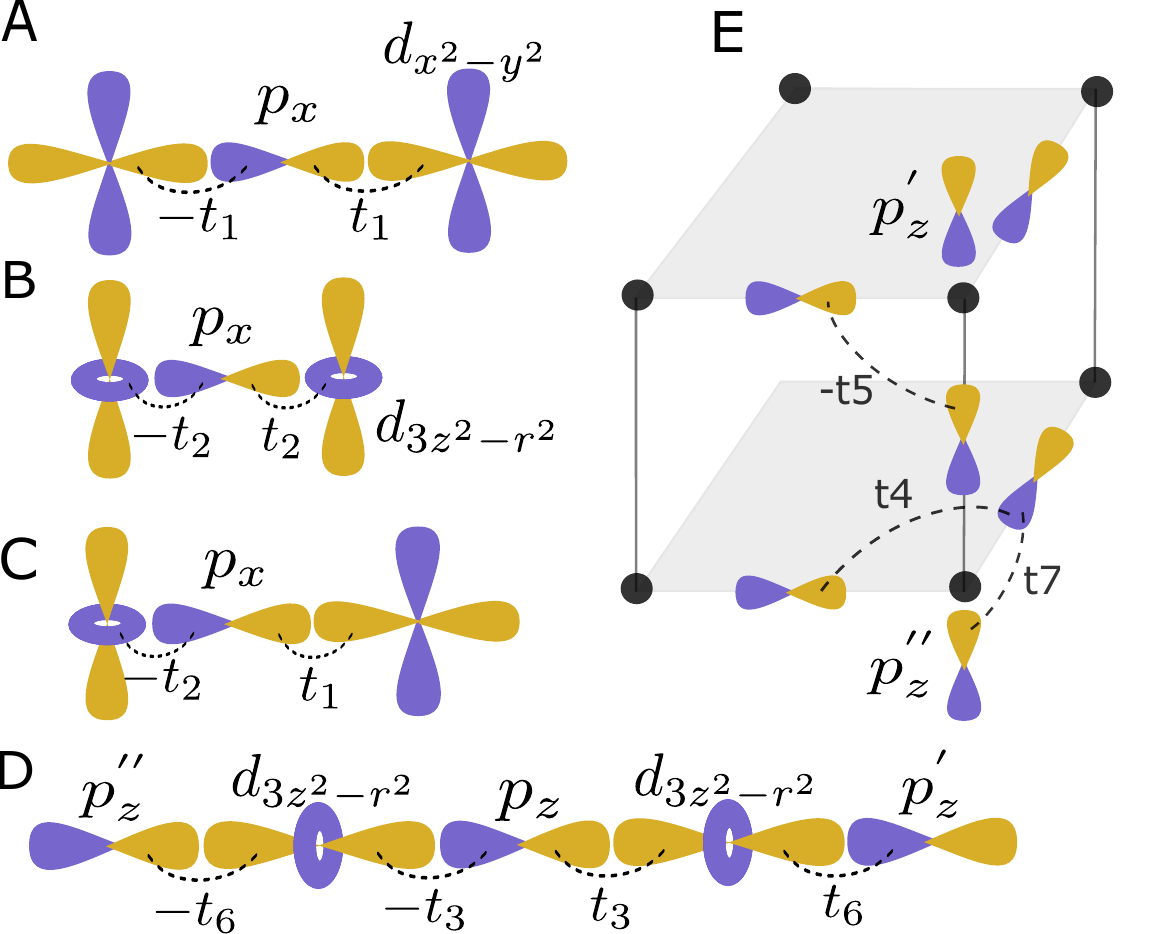}
\caption{ (Color online) The seven  hopping terms in our 11-band Hubbard model~\cite{luo2023bilayer}. 
\textbf{A-D:} Four hopping processes between Ni-$d$ and O-$p$ orbitals that  lead to major superexchanges between Ni-$d$ orbitals. Here $t_1 = -1.56, t_2 =0.75, t_3 = -1.63, t_6 =1.37$. The site energy $\epsilon_{\dx} = -1.06,\epsilon_{\dz} = -1.16, \epsilon_{p_x/ p_y} = -4.94 , \epsilon_{p_z} = -4.30, \epsilon_{p^{\prime \prime}_z} =\epsilon_{p^{\prime}_z}= -3.77 $.
Note that the hopping process \textbf(C) is from combining {A} and {B}. The  superexchange between the on-site intra-layer $\dx$ and $\dz$ orbitals (not shown here) vanishes due to symmetry.  \textbf{E:} Hoppings between O-$p$ orbitals  are shown  on a cartoon depicting the  structure of the bilayer $\mathrm{NiO_2}$ planes. The 3$d$ orbitals in $\mathrm{Ni}$ (black dots) are not shown for clarity. Here $t_4 = 0.58 , t_5 =0.49, t_7 =0.43$. These seven hopping terms combining the site-energies define our 11-band Hubbard model in the non-interacting limit.
} \label{fig:super}
\end{figure}


To address above questions, here we study an 11-band Hubbard model that includes four 3$\dx$ / 3$\dz$ orbitals of nickel,  and seven most relevant 2$p$ orbitals of oxygen in the $\mathrm{NiO_2}$ bi-layer per site.  We carry out determinant quantum Monte Carlo simulations (DQMC)~\cite{blankenbecler1981monte,assaad2008world} and cellular dynamical mean-field theory  (CDMFT)~\cite{georges96,maier05rmp} calculations  in the normal state of the system. 
Our result suggests that the  superexchange couplings in this system are in general comparable to  that in cuprates~\cite{weber2012scaling}, supporting a magnetic correlation origin of the high $T_c$  superconductivity in    $\mathrm{La_3Ni_2O_7}$  under pressure. We show that at large Hubbard $U$, the system is a charge-transfer insulator~\cite{karp2020} in the Zaanen-Sawatzky-Allen (ZSA) scheme~\cite{zsa1985} at half-filling ( \textit{i.e.}, four holes per site). Upon hole doping, the vertical Ni-$\dz$ - O-$\pz$ orbitals host a narrow low-energy spin-singlet band, where a small doping level is maintained over a broad range of hole doping of the system.  In contrast, the in-plane  Ni-$d_{x^2-y^2}$  and  $\mathrm{O}- \px / \py$ orbitals  form low-energy singlet band with high propensity for hole doping and greater itinerancy, drawing an analogy to the Zhang-Rice singlet band (ZRSB) in cuprates~\cite{zhang1988effective, kowalski2021oxygen,  shane2022}. 
 We discuss the implication of this inhomogeneous distribution of holes in the two $e_g$ orbitals to the superconductivity. An effective $t-J$ model that considers the leading-order exchange couplings in  $\mathrm{La_3Ni_2O_7}$ under pressure is proposed. The effects of the Hund's coupling is also discussed.
\section{Model and method}
To fully take into account the superexchange couplings, we consider an 11-band
Hubbard model~\cite{luo2023bilayer} that can be written as,
\begin{eqnarray}
H =  \sum_{i,j,\alpha,\beta,\sigma}t_{i,j,\alpha,\beta}d_{i \alpha \sigma}^{\dagger}c_{j \beta \sigma}+\sum_{i,j,\alpha,\beta,\sigma}t_{i,j,\alpha,\beta}c_{i \alpha \sigma}^{\dagger}c_{j \beta \sigma}    \nonumber  \\
 +\sum_{i \alpha \sigma}(\epsilon_{\alpha}-\mu)n_{i \alpha  \sigma}^{d} +\sum_{i \alpha \sigma}(\epsilon_{\alpha}-\mu)n_{i \alpha  \sigma}^{c}  -\sum_{i \alpha \sigma}E_{dc}n_{i \alpha \sigma}^{d}   \\ 
+\sum_{i \alpha}Un_{i \alpha \uparrow}^{d} n_{i \alpha \downarrow}^{d}+\sum_{i, \alpha < \beta,\sigma }U^{\prime}n_{i \alpha \sigma}^{d}n_{i \beta \bar{\sigma}}^{d}+ (U^{\prime} - J_H) n_{i \alpha \sigma}^{d}n_{i \beta \sigma}^{d}  \nonumber
\label{eq:hamil11}
\end{eqnarray}
where $t_{i,j,\alpha, \beta}$ denote hoppings between electrons on sites ($i,j$) and orbital ($\alpha, \beta$)( can be either Ni-$d$ or O-$p$ orbitals).  $d^{\dagger}_{\alpha,i,\sigma}$ ($c^{\dagger}_{\alpha,i,\sigma}$) is the creation operator for electrons on $\alpha \in $ 3$d$ ($\in$ 2$p$) orbital. $\epsilon_{\alpha}$  is  the site-energy of  $\alpha -$ orbital. $U$ is the Hubbard interaction between two electrons on the same $d$-orbital ($\dx$   or   $\dz$) and $U^{\prime}$ is for  that  on two different $d$-orbitals. $U^{\prime} = U-2J_{H}$ is adopted where $J_{H}$ is the Hund's coupling as following Ref.~\cite{held2001mott}.  $E_{dc}$ is the double counting (DC) term~\cite{anisimov1991,karolak2010double,wang2012coval} to be subtracted in the DQMC or CDMFT. Here we use the Held's formula~\cite{held2007electronic}: $E_{dc} = \frac{1}{3}(U+2U^{\prime}-J_H)(n^{0}_d -0.5) $, with $n^{0}_d $ being the occupation number of  $d-$orbitals in the non-interacting limit, $n^{0}_d \approx 2.16$ in our case (see also the \textit{Appendix}).
 We adopt the hopping parameters and site-energies  proposed in Ref.~\cite{luo2023bilayer}, which is obtained by downfolding the DFT result in the maximally localized Wannier orbitals. See Fig.~\ref{fig:super} and  Fig. S1 in SI Appendix for all the hopping parameters and site-energy values. In line with DFT result~\cite{luo2023bilayer},  we assume that  the chemical potential $\mu = 0$ in above Hamiltonian corresponds to the pristine single crystal $\mathrm{La_3Ni_2O_7}$  material at pressures $> 14$GPa,  without considering other  potential doping effects (oxygen deficiency for example). We will also vary $\mu$ to explore regimes with different hole concentrations that defined  as  $n_h = ( 22- \sum_{\alpha , \sigma} n_{\alpha, \sigma})/ 4 $, namely, the average number of holes per $d-$ orbitals per site. $n_h = 1 $ corresponds to half-filling in our study.  In this work, we use $U=7$ eV and $J_H = 0.1U = 0.7$ eV unless otherwise stated.  Below we use electron volt ($\mathrm{eV}$) as the energy unit throughout the paper. For the calculations on cuprate, we employ a canonical set of parameters for the three-band Hubbard model~\cite{weber2012scaling} (one $\dx$ orbital and two $\px$ / $\py$ orbitals in the $\mathrm{CuO_2}$ plane):  $t_{pd} = 1.39, t_{pp} = 0.64, t^{\prime}_{pp} = 0.103, \Delta_{dp} \equiv \epsilon_d - \epsilon_p = 2.6, U = 8.5, \mathrm{E_{DC}} = 3.12$. This set of parameters is assumed to be most relevant for the $\mathrm{La_{2-x}Sr_xCuO_{4}}$ (LSCO) compound, which has been used in   studies on different aspects of cuprates~\cite{weber2012scaling,kung2016characterizing,kowalski2021oxygen,mai2021pairing}.

\begin{figure}[h]
\includegraphics[scale=1.0 ]{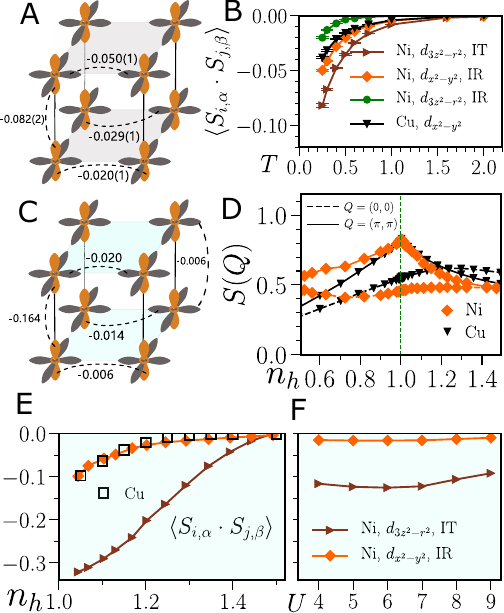}
\caption{Magnetic correlations between $d$-orbitals due to  the superexchanges in the 11-band Hubbard model. \textbf{A,C:} The spin-spin correlation function $\langle S_{i,\alpha} \cdot S_{j, \beta} \rangle $ for four neighboring d-orbitals are shown  in numbers to profile the relative strength of the antiferromagnetic  superexchange couplings in the system. Brown (Gray) symbols denote $\dz$ ($\dx$) orbitals.  \textbf{A} and \textbf{C} are respectively from  DQMC at half-filling ($n_h = 1$), $T=0.25$, and from CDMFT at $\mu =0$ ($n_h \approx 1.244$),$T=0.08$ . Note that here the magnetic correlations between the on-site $\dx$-$\dz$ orbitals due to Hund's coupling are not shown.  \textbf{B:} $\langle S_{i,\alpha} \cdot S_{j, \beta} \rangle $ between  pairs of nearest neighbouring (NN) $d$ orbitals in Ni-11-band and Cu-3-band model as a function of temperature $T$ at half-filling from DQMC.
Diamonds show result for intra-layer (IR)  $\dx$- $\dx$ correlations, while triangles denote that between the inter-layer (IT) $\dz$ - $\dz$ orbitals.
\textbf{D:} The spin structure factor $S(Q) = \frac{1}{N} \sum_{i,j} \langle  S_{i\alpha} \cdot S_{j\beta} \rangle e^{-iQ \cdot (R_i - Rj)}$ for Ni (intra-layer, \textit{i.e.}, $\alpha = \beta = \dx $ component) as a function of hole concentration $n_h$  compared with that of cuprate. \textbf{E:} $\langle S_{i,\alpha} \cdot S_{j, \beta} \rangle $ as a function of $n_h$ from CDMFT at $T=0.08$.  \textbf{F:} Same as E but result is shown as a function of Hubbard $U$ ($J_H = 0.1U$) at $n_h=1.25 (\mu \sim 0 )$.
 Here $U = 7, J_H = 1.05$ for nickelate and $U = 8.5$ for cuprate unless otherwise stated. In DQMC, we use $6 \times 6 \times 11$ orbitals for $\mathrm{Ni}$ and  $6 \times 6 \times 3$ orbitals for Cu. } 
\label{fig:mag}
\end{figure}

For DQMC simulation, we employ two dimensional square lattices with $6 \times 6 \times 11=396$ orbitals at most. For CDMFT study, we carry out computations in the normal state, where an effective impurity model with the $ 2 \times 2\times 11=44$ orbitals is used. For more details on the methods see the \textit{Appendix}.
 
\section{Results}

\subsection{Superexchanges} We first discuss the  magnetic exchange couplings in the system. As shown in Fig.~\ref{fig:super}, there are a few hopping processes can give rise to significant superexchanges. In the atomic limit of the charge-transfer picture, the spin singlet state of two Ni-$d$ electrons acquires an energy gain of $J = \frac{-4t^{4}_{pd}}{(U+\Delta_{pd})^2 }\times (\frac{1}{U}+\frac{1}{U+\Delta_{pd}})$ over the spin triplet states, where $\Delta_{pd} = \epsilon_d - \epsilon_p$, and $t_{pd}$ is the hopping between Ni-$d$ and O-$p$ orbitals.
  The numbers shown in Fig.~\ref{fig:mag}A are  the magnitudes of spin correlation function $\langle S_{i,\alpha} \cdot S_{j, \beta} \rangle $ for a few pairs of neighbouring $d-$ orbitals at $T=0.25$ from DQMC. Here $ S_{i,\alpha} $ is the spin operator at site $i$ and orbital $\alpha$. This  result profiles the relative strengths of the main exchange couplings in the system: at $T=0.25$ and half-filling,  the inter-layer (IT) on-site $\dz$-$\dz$ antiferromagnetic (AFM) exchange dominates the  exchange couplings in the system [ $ \langle S \cdot S \rangle=-0.082(2)$]. Then it comes the intra-layer (IR) nearest-neighboring  $\dx$-$\dx$
exchange [$ \langle S \cdot S \rangle=-0.050(1)$]. The intra-layer  $\dz$-$\dx$ superexchange is weaker than the aforementioned  two [$ \langle S \cdot S \rangle=-0.029(1)$]. 
Finally, the intra-layer $\dz$-$\dz$  exchange is found the weakest among those couplings. In particular, in CDMFT at $T=0.08$ and   $\mu=0$ (Fig.~\ref{fig:mag}C),  the intra-layer $\dz$-$\dz$  correlation is less than 1/20 of the inter-layer $\dz$-$\dz$ correlation, hence it can be neglected in future studies on  $\mathrm{La_3Ni_2O_7}$ ( which corresponds to $\mu=0$ in our study). 
Fig.~\ref{fig:mag}B and~\ref{fig:mag}D show  $\langle S_{i,\alpha} \cdot S_{j, \beta} \rangle $ as a function of temperature at half-filling, and spin structure factor $ S_{\alpha,\beta}(Q) $ as a function of $n_h $ at $T=0.3$ respectively,  comparing with the result of the 3-band Hubbard model of  cuprates.  As one can see that the intra-layer $\dx$-$\dx$  correlation in  the 11-band Hubbard model  is  in general comparable to its cuprate counterpart. The inter-layer $\dz$-$\dz$ correlation, on the other  hand, seems to be essentially  stronger  the $\dx$-$\dx$  correlations. This result implies that the AFM correlations between inter-layer $\dz$ orbitals could be at the origin of the observed superconductivity in $\mathrm{La_3Ni_2O_7}$ under pressure. Fig.~\ref{fig:mag}E and~\ref{fig:mag}F show respectively $\langle S_{i,\alpha} \cdot S_{j, \beta} \rangle $  as a function of $n_h$ at $U = 7$, and as a function of $U$ at $n_h = 1.25 $ (where $\mu \sim 0$ )  in CDMFT. Here one sees that varying the value of $U$ in the range of $(4 \sim 9)$ eV does not change substantially the interlayer $\dz$-$\dz$ magnetic correlation at $\mu \sim 0$ (Fig.~\ref{fig:mag}F). More importantly, as shown in Fig.~\ref{fig:mag}E, the interlayer $\dz$-$\dz$ AFM correlation does not vanish unless a huge hole doping $p = n_h-1 \gtrsim 0.45$ is approached.  We find that despite the fact that Hund's coupling $J_H$  in principle can transfer the magnetic correlations between the vertical $\dz - \dz$ bonds to other orbitals, the overall effects seem to be negligible in our study. For example, as shown in Fig.\ref{fig:mag}C,  $\langle S \cdot S \rangle$ between interlayer $\dx - \dx$ orbitals are tiny ($\sim -0.006$) as compared to that of the  $\dz - \dz$ orbitals. This may be attributed to the small effective filling factors of the in-plane orbitals, namely, the in-plane $\dx$ and $\px / \py$ orbitals are less affected  by the Hund's coupling due to their large itinerancy, as to be shown bellow.

\begin{figure}[h]
\includegraphics[scale=0.6]{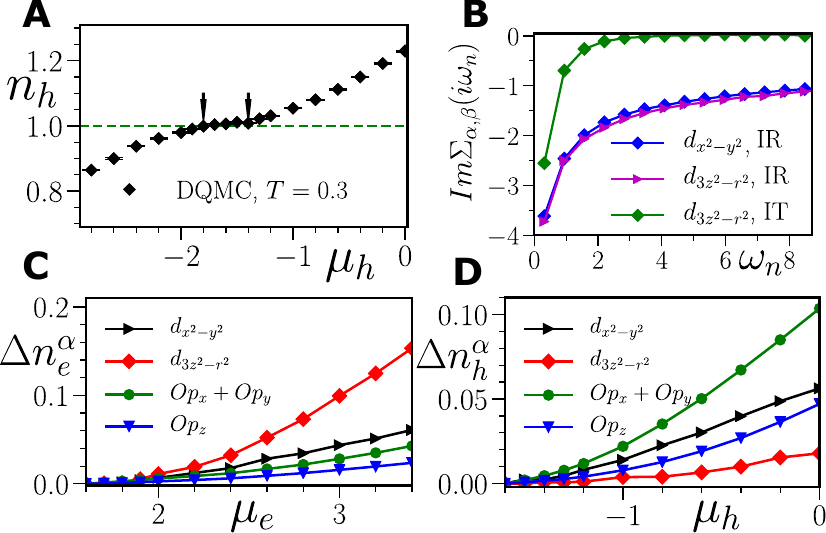}

\caption{ Charge transfer insulating behaviour of the 11-band Hubbard at half-filling. \textbf{A:} The hole concentration $n_h$ as a function of hole chemical potential $\mu_h$. DQMC result of $n_h$ at $T=0.3$ suggests that a charge gap opens as hole chemical potential approaches $\mu_h \sim -1.6$. Black arrows indicate the $\mu_h$ range where $n_h \approx 1$.  \textbf{B:} Imaginary part of the self-energies from DMFT at  half-filling $n_h \approx 1$ and $T=0.1$. \textbf{C:} The change of electron concentration $\Delta n^{\alpha}_e$  for different orbital-$\alpha$ as a function of electron chemical potential $\mu_e$ for  $\mu_e > 1.6$ (where $n_e > 1$, electron-doped). $\Delta n^{\alpha}_e$ is defined as  $\Delta n^{\alpha}_e  \equiv n^{\alpha}_e (\mu_e) -n^{\alpha}_e (\mu_e = 1.6) $, where $n^{\alpha}_e (\mu_e = 1.6) \approx 1$ is the electron filling of orbital-$\alpha$ at  $\mu_e =-\mu_h = 1.6$ (  $ n_e = \sum_{\alpha}n^{\alpha}_e $ ). \textbf{D:} Result similar to C but for  $\Delta n^{\alpha}_h$ in hole doing regime ($\mu_h > -1.6$, $n_h = 2 -n_e > 1$) . C and D use same method and same parameters as A. See also main text.
}
\label{fig:cti}
\end{figure}

\subsection{Charge transfer insulator} We now focus on the metal- insulator transition in the 11-band Hubbard model. Fig.~\ref{fig:cti}A displays the hole concentration per $d$ orbital $n_h$ as a function of hole chemical potential $\mu_h \equiv - \mu$. As one can see that the DQMC result  (dots)  of $n_h$ at $T=0.3$ exhibits almost a plateau at $n_h  =1$ ( $\mu_h \sim -1.6$) with  a vanishing compressibility $  \partial n_h / \partial \mu_h$, suggesting the opening of a charge gap at half-filling.  CDMFT result on the imaginary part of the self-energies obtained at lower temperature shown in Fig.~\ref{fig:cti}B also points to the insulating nature of the 11-band Hubbard model at half-filling. To understand the nature of this insulating state, we  further study how the hole/electron concentration changes in different orbitals as a function of chemical potentials in Fig.~\ref{fig:cti}C and Fig.~\ref{fig:cti}D respectively. In other words, as hole chemical potential $\mu_h \equiv -\mu$ and electron chemical potential $\mu_e \equiv \mu$ are increased from their half-filled values $ \mu_h = -\mu_e \sim -1.6$, holes/electrons can be added into different Ni-$d$ and O-$p$ orbitals of the system, which is denoted by $\Delta n^{\alpha}_{e/h}$ in Fig.~\ref{fig:cti}C and  Fig.~\ref{fig:cti}D. As one can see that the doped holes go  primarily to the oxygen $p-$ orbitals (Fig.~\ref{fig:cti}D), while the doped electrons reside mainly on the nickel $d-$ orbitals (Fig.~\ref{fig:cti}C),  which unambiguously indicates the charge-transfer nature~\cite{zsa1985} of  the insulating state  of this system at half-filling. Similar to cuprates~\cite{kowalski2021oxygen},  here $\dx$ orbital  has a sizable portion of doped holes in the case of hole doping ($\mu_h > -1.6$). It is remarkable that, as shown in Fig.~\ref{fig:cti}D, $\Delta n^{\alpha}_{h}$ of $\dz$ orbital are much smaller than that of $\dx$ orbital in the hole doped regime at given $\mu_h$, reflecting the fact the $\dz$ orbital is more strongly correlated than the $\dx$ orbital at hole doping. See more discussions below.  We note that in cuprates,  a smaller portion of holes residing on cations in general indicates  a larger superexchange coupling~\cite{ruan2016relationship,kowalski2021oxygen}.

\subsection{Zhang-Rice singlet band}  
 In Fig.~\ref{fig:akw}A, we plot the local density of states (DOS) $\rho(\omega)$ for Ni-$\dz$,  and for the out-of-plane O-$\pz$ orbital at $n_h \approx 1.035, T=0.08$, where we see that the overall structure of $\rho(\omega)$ resembles that of the  underdoped cuprate~\cite{kowalski2021oxygen}: the well-separated lower and upper Hubbard bands (LHB/UHB) appear in the Ni DOS. There is a central band with mixed weights of Ni-$d$ and O-$p$ orbitals that can be called the charge-transfer band(CTB)~\cite{kowalski2021oxygen}. More importantly, the DOS of  Ni-$\dz$ and O-$\pz$ near Fermi level form a narrow band separated from the upper Hubbard band by a charge-transfer gap (CTG). In the context of cuprate, this  low-energy band is usually referred to as the Zhang-Rice singlet band (ZRSB)~\cite{zhang1988effective}. Similar result of the in-plane Ni-$\dx$ and  O-$\px / \py$ orbitals is shown in Fig.~\ref{fig:akw}B, where the low-energy singlet band has a wider bandwidth compared to its out-of-plane  counterpart shown in Fig.~\ref{fig:akw}A, suggesting a less localized nature of the  in-plane orbitals.
 

There are a few points we would like to emphasize. First, the vertical $\dz$-$\pz$ singlet band can be quite different from a conventional ZRSB. In cuprates,  a doped hole at oxygen sites hybridize with a $\mathrm{Cu^{2+}}$  hole in terms of the superposition of  four O$p$ hole states  adjacent to the $\mathrm{Cu^{2+}}$ iron, forming a spin singlet. The spin singlet moves effectively in the antiferromagnetic background of $\mathrm{Cu^{2+}}$ lattice with an effective bandwidth being  $(2 \sim 3)$eV~\cite{kowalski2021oxygen}.  Here the vertical singlet states of Ni-$\dz$-O-$\pz$ have a narrow bandwidth, and they barely  interact with each other directly. Instead, the in-plane holes hybridize with the vertical singlets via $\dz$ - $\px$/$\py$ hopping ($t_2 = 0.75$), or via the $\px$ -$\pz$ hopping ($t_5=0.49, t_7=0.43$ , see  Fig.~\ref{fig:super}). As a result,  the vertical $\dz$ singlet band may behave more like scattering centers  with antiferromagntic characteristics in the system.
On the other hand, one can expect that the in-plane ZRSB associated with the $\dx$ and $\px / \py$ orbitals could draw a close analogue to the ZRSB in cuprate. However, as exemplified in Fig.~\ref{fig:akw}D for $n_h \approx 1.244$ ($\mu=0$), the in-plane orbitals become itinerant and the charge-transfer gap is essentially absent at this large hole concentration.  Hence, whether the concept of ZRSB still applies in capturing the in-plane single particle excitations in $\mathrm{La_3Ni_2O_7}$  is in question.  It is remarkable that, nevertheless, the vertical singlet band as shown in Fig.~\ref{fig:akw}C ,  remains intact at this large $n_h \approx 1.244$. 

\begin{figure}[h]
\includegraphics[scale=0.8 ]{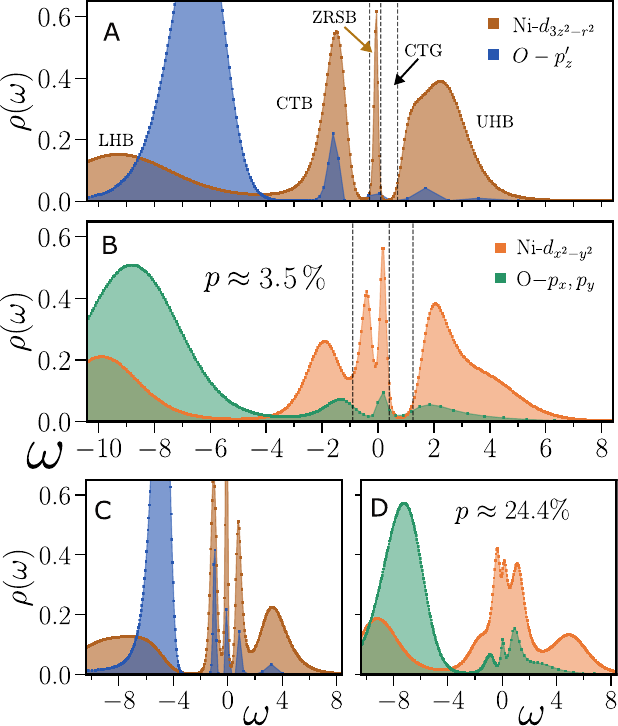}
\caption{Local density of states (DOS). \textbf{A:} DOS  of the out-of-plane $\dz$ and $p^{'}_z$ orbitals. Here $n_h =\approx 1.045$ hence $p=n_h-1 \approx 3.5\%$
\textbf{B:} DOS  of the in-plane $\dx$ and $\px$ ($\py$) orbitals.  Dashed vertical lines indicate the Zhang-Rice singlet band (ZRSB) and the 
charge-transfer gap (CTG). UHB stands for upper Hubbard band, and LHB stands for lower Hubbard band. \textbf{C:} The same as A but $p \approx 24.4\%$ ($\mu=0$).  \textbf{D:} The same as B but $p \approx 24.4\%$.
 DOS are obtained by maximum entropy (MEM) analytic continuation~\cite{dominic16} of the Matsubara Green's functions that obtained by CDMFT at  $T=0.08, U=7, U' =0.8U, J_H = 0.1U$. Similar results can be seen in  Fig. S2 in Appendix at different Hubbard $U$ and temperature $T$.
}
\label{fig:akw}
\end{figure}

\begin{figure}[h]
\includegraphics[scale=0.6 ]{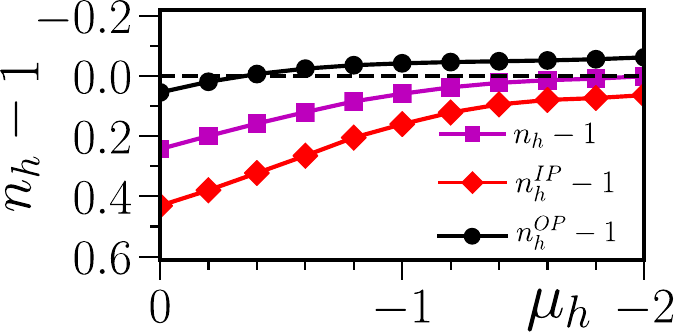}
\caption{ Hole concentrations as functions of chemical potential $\mu_h$.
Here  $n^{IP}_{h} = n^{h}_{\dx} + n^{h}_{\px} + n^{h}_{\py}$ counts the number of holes in the  in-plane $\dx$ orbital and $\px$ /$\py$ orbitals combined, while $n^{OP}_{h} = n^{h}_{ \dz} + 0.5(n^{h}_{\pz} + n^{h}_{\pz^{\prime}}+n^{h}_{\pz^{\prime\prime})}$ counting the number of holes in the out-of-plane  orbitals. $n_h =  (n^{IP}_{h}+  n^{OP}_{h })/2$. Thus  $n^{h}-1$ is the hole doping from half-filling of the system.  Note that as $\mu_h$ approaches $\mu_h=-2$, $n_h-1$ curve becomes flat, indicating the opening of a charge gap at half-filling.
}
\label{fig:dist}
\end{figure}

\begin{table}[h]
\begin{tabular}{ccccc}
\hline \hline 

   & $\hspace{0.2cm} n_{h}$ & $\hspace{0.8cm}$ $n^{IP}_{h} \hspace{0.9cm} $ & $ n^{OP}_{h}$ \tabularnewline\hline 

 DQMC ($U=7, T=0.25$) & 1.233 & 1.400 & 1.065 \tabularnewline \hline 
 CDMFT ($U=7,T=0.08$)  & 1.244 &  1.434 & 1.054 \tabularnewline \hline
 CDMFT ($U=7^{*},T=0.1$)  & 1.256 &  1.443 & 1.069\tabularnewline \hline
 CDMFT ($U=9 ^{*},T=0.1$)  & 1.216 &  1.385 & 1.048 \tabularnewline \hline \hline

\end{tabular}
\caption{\label{tab:hop1} Hole concentrations at $\mu_h =-\mu = 0$. The last two results starred out are obtained with $U^{\prime} = 0.7U, J_H=0$, \textit{i.e.,} there is no Hund's coupling.} 
\label{tab:1}
\end{table}

 Now we focus on  the distribution of holes in different orbitals. As discussed above, considering the formation out-of-plane ( in-plane ) spin singlet states, it is judicious to consider the carriers in the $\dz$ and $\pz$ (  $\dx$ and $\px / \py$)  orbitals  as a whole, which are represented by the out-of-plane hole concentration $n^{OP}_h$ ( in-plane hole concentration $n^{IP}_h$) in Fig.~\ref{fig:dist}.  In Fig.~\ref{fig:dist}, we see that the out-of-plane  hole doping level $p = n^{OP}_h -1$ (dots) is always small while $\mu_h$ is varied in the range of ($ 0 , -2$). In other words, the  out-of-plane orbitals as a whole is always underdoped ( $|n^{OP}_h -1| < 7\%$) in this chemical potential range.  The in-plane orbitals, in contrast, can be heavily overdoped, as denoted by a large value of $(n^{IP}_h -1)$ (diamonds) when $\mu_h \sim 0$.  In Table-I we inspect specifically the  hole concentration levels at $\mu =0$, which is assumed to correspond to that of the real material of $\mathrm{La_3Ni_2O_7}$ under pressure. As one sees that, in our study the 11-band Hubbard model  is about  $ p= 22 \sim 26\%$ hole  doped at $\mu =0 $. Changing methods or varying the values  of Hubbard $U$, Hund's coupling $J_H$ or temperatures $T$ do not  significantly change the value  of $p$. This result roughly coincides with the nominal doping level of the $e_g$ orbitals in $\mathrm{La_3Ni_2O_7}$  $( p=25\%)$.  Table-I also shows that at $\mu = 0$, the hole doping in the out-of-plane orbitals is around $p = (n^{OP}_h - 1) \approx ( 5\% \sim 7 \%)$ , while for  that of the in-plane orbitals,  $p = (n^{IP}_h -1 )$  is about $\approx  40\%$ at different Hubbard $U$ and temperature $T$ . The stark inhomogeneous distribution of the holes in the two different $e_g$ orbitals ( and their correlated O-$p$ orbitals ) explains the finding in Fig.~\ref{fig:akw} that the vertical  low-energy singlet band remains strongly correlated at large $p$ while the in-plane low-energy DOS displays great itinerancy.
A natural question is that whether the in-plane $\dx$ and $\px / \py$ orbitals can be seen as pure  itinerant orbitals  without interaction terms in $\mathrm{La_3Ni_2O_7}$? 
We find that in CDMFT at $n_h \sim 1.25$, the magnetic correlation between intra-layer $\dx -\dx$  orbitals is weak but not vanishing ( see Fig.~\ref{fig:mag}E).  It is  notable that the strange metal (SM) phase can be very sensitive to weak magnetic correlations~\cite{wu2022}. Hence, if the observed strange metal state in  $\mathrm{La_3Ni_2O_7}$ at $P > 18 \mathrm{GPa}$ is  also  related to the magnetic correlations in  $\dx$ orbitals, then the latter may not be simplified as pure itinerant.

\subsection{Effective t-J model} Based on our study above, we propose a four-band $t-J$ model to describe the low-energy physics of the $\mathrm{La_3Ni_2O_7}$ under pressure. The proposed Hamiltonian can be written as $\mathcal{H} = \mathcal{H}_{0}+\mathcal{H}_J$, with the non-interacting $\mathcal{H}_{0}=\sum_{{\rm k}\sigma}\Psi_{{\rm k}\sigma}^{\dagger}H({\rm k})\Psi_{{\rm k}\sigma} $ written as,

\begin{eqnarray}
\label{eq:H0}
H(k)_{1,1} = H(k)_{3,3}  = -2t^x_1 [\rm cos(k_x) +cos(k_y)] \nonumber  \\
    - 4t^x_2 \rm cos(k_x)cos(k_y) + \epsilon_x \nonumber \\
H(k)_{2,2} = H(k)_{4,4}    = -2t^z_1 [\rm cos(k_x) +cos(k_y)] \nonumber \\
      - 4t^z_2 \rm cos(k_x)cos(k_y)+ \epsilon_z \nonumber \\
      H(k)_{1,2} =  H(k)_{3,4}   = -2t^{xz} [\rm cos(k_x) -cos(k_y)]  \nonumber  \\
 H(k)_{2,4}    = -2V_{\perp} [\rm cos(k_x) -cos(k_y)] \bigskip   
 \end{eqnarray}
 
  where $\Psi_{\sigma}=\left(d_{x_1 \sigma},d_{z_1\sigma},d_{x_1\sigma},d_{z_2\sigma}\right)^{T}$, denoting the annihilation operators of $\dx$ and $\dz$ orbitals in the two $\mathrm{NiO_2}$ layer, while the oxygen degrees of freedom have been integrated out.   The $ \mathcal{H}_{0}$  part is taken from the down-folded tight-binding model from Luo \textit{et al}'s work in Ref.~\cite{luo2023bilayer}, namely, 
 $t^x_1 \approx 0.5 , t^x_2 \approx 0.07, t^z_1 \approx 0.11, t^x_2 \approx  0.02, t^{xz} = -0.24  , V_{\perp} = 0.64$ .
 
 For the interacting part $\mathcal{H}_{J}$,  we consider three main magnetic exchanges terms,
 \begin{eqnarray}
\mathcal{H}_{J} =  J_1 \sum_{ i  }   { ( S_{i, z_1} S_{i, z_2} -\frac{1}{4} n_{i,z_1}n_{i,z_2})}   \nonumber   \\
  + J_2 \sum_{ \langle i,j\rangle , \alpha = x_1,x_2  }  ( S_{i, \alpha} S_{j, \alpha} -\frac{1}{4} n_{i, \alpha}n_{j, \alpha} ) \nonumber\\
  + J_3 \sum_{ \substack{  { \langle i,j\rangle} \\  {( \alpha,\beta) = (x_1, z_1) / (x_2 ,z_2)} }   }   ( S_{i, \alpha} S_{j, \beta} -\frac{1}{4} n_{i, \alpha}n_{j, \beta} ) \label{eq:Hj}
\end{eqnarray}

where $J_1$ captures the exchange couplings between the on-site inter-layer  $\dz$ orbitals, $J_2$ the exchanges between the intra-layer $\dx$   orbitals on  nearest neighboring (NN)  sites, and finally $J_3$ for the intra-layer  $\dx$ -  $\dz$ exchanges on  NN sites. Considering the superexchange coupling calculated in the atomic limit with $U=7$, and the magnetic correlation results compared with LSCO cuprate~\cite{wakimoto2007, wang2022paramagnons}, typical value of $J_1$ can be chosen as, for example,  $J_1 \sim 0.18 \mathrm{eV}$ (see \textit{Appendix}). The values of $J_2, J_3$ should be in principle smaller than $J_1$. The arbitrariness of adopting the values of these $Js$, however, can not be eliminated without further experimental researches. The magnetic couplings between the on-site interlayer $\dx - \dz$ and interlayer $\dx - \dx$ orbitals may also be further added in Eq.~\ref{eq:Hj} when Hund's coupling $J_H $~\cite{medici2011janus} is taken into account.

\section{Discussion and Conclusion}  After the discovery  of high-$T_c$ superconductivity in  $\mathrm{La_3Ni_2O_7}$ under pressure~\cite{sun2023},  a number of effective interacting models have been proposed to study the pairing symmetry~\cite{yang2023possible,shen2023effective,gu2023effective,sakakibara2023possible}, as inspired by the \textit{ab initio} calculations~\cite{luo2023bilayer,zhang2023electronic,lechermann2023electronic,christiansson2023correlated,shilenko2023correlated} or phenomenological insights. These effective models ignore the oxygen degrees of freedom, incorporating the magnetic correlations between Ni orbitals in terms of direct exchanges. Hence, the charge transfer property, which has shown to be one of the key  ingredients in determining $T_c$ of  cuprate superconductors~\cite{weber2012scaling,ruan2016relationship, rybicki2016perspective, kowalski2021oxygen,Mahony2022electron},  as well as the superexchange couplings, are beyond the scope of those effective models. In this work, we have studied an 11-band Hubbard model including both Ni-$3d$ orbitals and relevant O-$2p$ orbitals. We reveal the relative strengths of the superexchange couplings between different Ni-$3d$ orbitals. We find two spin-singlet bands in the system possessing strikingly different  hole concentrations and correlation strengths. These results suggest that the strong antiferromagnetic correlations within the inter-layer $\dz$ orbitals might be the driving force of the SC in $\mathrm{La_3Ni_2O_7}$. The role of the more itinerant $\dx$ orbital in SM and SC states, owing to the small but not vanishing remnant magnetic correlations within, and the specific geometry of the $\alpha , \beta -$ sheets of the Fermi surface~\cite{luo2023bilayer}, might be complicated and requires further investigations. Finally, we note that an experimental work~\cite{Liu2023wen} probing the optical response in  $\mathrm{La_3Ni_2O_7}$ reveals the  proximity to Mottness of the electron correlations. This finding is in accordance with our doped charge-transfer insulator description of  the $\mathrm{La_3Ni_2O_7}$ under pressure.
 

\section{Acknowledgements}
We thank Hualei Sun, Mi Jiang, K. Le Hur, Yi Lu, and Xunwu Hu for useful discussions. W.W is  indebted to A. -M. Tremblay for useful discussions and  insightful suggestions. W.W. acknowledge help from Dong Meng in preparing the illustrations in Figure 1.  Work at Sun Yat-sen University was supported by the National Natural Science Foundation of China (Grants  No.12274472, No. 92165204, No.12174454, No.11974432), the National Key Research and Development Program of China (Grants No. 2022YFA1402802, 2018YFA0306001),  the Guangdong Basic and Applied Basic Research Foundation (Grants No. 2022A1515011618, No. 2021B1515120015), Guangdong Provincial Key Laboratory of Magnetoelectric Physics and Devices (Grant No. 2022B1212010008), Shenzhen International Quantum Academy (Grant No. SIQA202102), and Leading Talent Program of Guangdong Special Projects (201626003). We
acknowledge the support from GuangZhou National Supercomputing Center (Tianhe-II).


\bibliography{Ni}

\newpage

\section{Appendix}

\newcommand{\ssin}[1]{\mathrm{sin(#1)}}

\renewcommand{\theequation}{S\arabic{equation}}
\renewcommand{\thefigure}{S\arabic{figure}}
\renewcommand{\thetable}{S\arabic{table}}

\section{Methods}

 For DQMC simulation, we use a two dimensional $6 \times 6 \times 11=396$ orbitals square lattice with periodic conditions for the 11-band Hubbard model, on which we have verified that the finite size effects are negligible in the parameter regime we study (See SI Appendix, Fig. S3). In DQMC we use  $\Delta \tau = 0.0625$.  For CDMFT study, we carry out computations in the normal state, where the $ 2 \times 2\times 11=44$ orbitals cluster  effective impurity model is used. At high temperatures, we find excellent agreement between the DQMC and CDMFT result(see Fig.~\ref{fig:cmp}).  The Hirsch-Fye quantum Monte Carlo (HFQMC) is used as impurity solver in CDMFT, where  $\Delta \tau = 0.078$ is adopted.

We note that in our study, due to the fact that $n^{0}_d $ is close to $2$, different DC schemes in fact give similar $E_{dc}$, hence lead to similar result. For example, at $\mu=0$,  using the Held's $E_{dc}$ we get $n_h \approx 1.24, n^{IP}_h \approx 1.43,  n^{OP}_h \approx 1.05 $, while the fully localized or atomic limit (FLL) $E_{dc}$ gives $n_h \approx 1.21, n^{IP}_h \approx 1.40,  n^{OP}_h = 1.02 $.  We have also carefully checked that the DC term we use does not significantly shift the non-interacting Fermi surface.

\subsection{The eleven-band Hubbard model} \label{Sup:Geometry}

The tight-binding Hamiltonian $H_0$ of our  11-band Hubbard model  is obtained from the Wannier downfolding of the DFT band structure as done in Ref.~\cite{luo2023bilayer}. The values of the  hopping amplitude and site-energies of $H_0$  are explicitly shown in the main text. The non-zero hoppings between $\alpha-$, $\beta-$ orbitals [ $H_0(\alpha, \beta )$ ] can be listed as below (for $\alpha<\beta$),
\begin{widetext}
\begin{eqnarray}
H_0(1,5)=H_0(3,7) = -2i*t_1\ssin{0.5k_x} &  H_0(1,6) =H_0(3,8) = 2i*t_1\ssin{0.5k_y}  \nonumber  \\
H_0(2,5) =H_0(4,7)= 2i*t_2\ssin{0.5k_x} & H_0(2,6) =H_0(4,8)= -2i*t_2\ssin{0.5k_x}   \nonumber \\
H_0(2,9) = t_3  & H_0(4,9) = -t_3 \nonumber \\
& H_0(5,6) = H_0(7,8 )= -4*t_4\ssin{0.5k_x}\ \ssin{0.5k_y} \nonumber \\
H_0(5,9) =  -2i*t_5\ssin{0.5k_x} & H_0(6,9) = -2i*t_5\ssin{0.5k_y} \nonumber \\
H_0(7,9) =  2i*t_5\ssin{0.5k_x} & H_0(8,9) = 2i*t_5\ssin{0.5k_y} \nonumber \\
H_0(2,10) = t_6 &   H(4,11) = -t_6 \nonumber \\
H_0(5,10) = 2i*t_7\ssin{0.5k_x} & H_0(6,10) = 2i*t_7\ssin{0.5k_y}  \nonumber \\
H_0(7,11) = -2i*t_7\ssin{0.5k_x} & H_0(8,11) = -2i*t_7\ssin{0.5k_y}  
\end{eqnarray}
\end{widetext}
where $t_1 = -1.56, t_2 =0.75, t_3 = -1.63,t_4 = 0.58,t_5 =0.49, t_6 =1.37, t_7 =0.43$. The $H_0(\beta, \alpha )$ can be obtained as the hermitian conjugate of $H_0(\alpha, \beta )$. The site-energies read,
\begin{widetext}
\begin{eqnarray}
H_0(1,1) = H_0(3,3) = -1.06 & H_0(2,2) = H_0(4,4) =-1.16 \nonumber \\
 & H_0(5,5) = H_0(6,6) = H_0(7,7) =H_0(8,8)= -4.94  \nonumber \\
  H_0(9,9) = -4.30 & H_0(10,10) = H_0(11,11) =-3.77 
\end{eqnarray}
\end{widetext}

\begin{figure}
  \begin{center}
    \includegraphics[width=0.9\columnwidth]{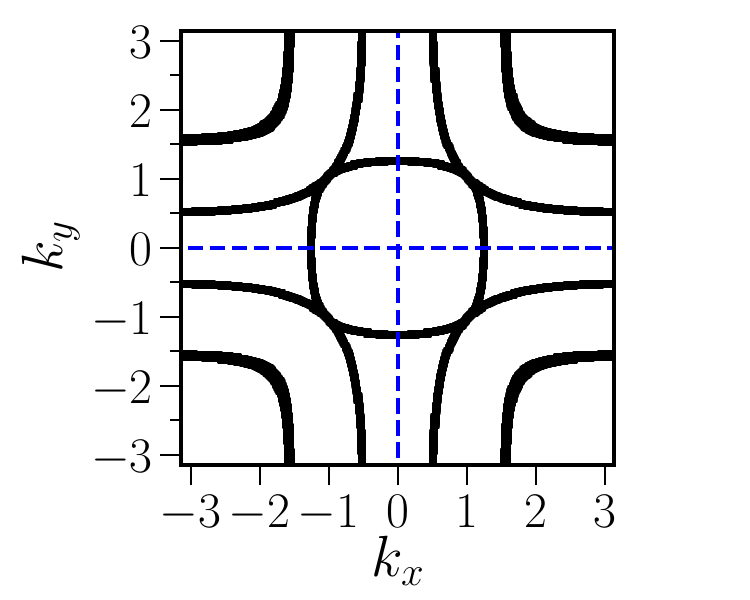}
  \end{center}
  \caption{
    Fermi surface of our 11-band tight-binding model. 
    }
\label{fig:fs}    
\end{figure}

The Fermi surface of our Hamiltonian $H_0$ described  above can be find in Fig.~\ref{fig:fs}, which reproduces the Fermi surface found in DFT computations~\cite{luo2023bilayer} on $\mathrm{La_3Ni_2O_7}$ under high pressures.

For the interaction part of  the 11-band Hubbard model, the Hund's coupling is adopted following Ref.~\cite{held2001mott}, namely, we take into account the longitudinal
component and neglect the transverse components of the Hund’s couplings. It has been shown that this approximation  affect  the Mott physics only  in a very minor way, as compared to the more rigorous  “Kanamori” Hund' coupling ~\cite{medici, quan2017competing}.  In Table-I in the main text, we have seen that even fully suppressing the Hund's coupling $J_H$, the result on the  distribution of the holes does not significantly change. 

Below we present a figure showing the local density of state (DOS) at $U=9, U^{\prime} = 6.3, J_H = 0,  n_h=1.05, T=0.1$, from which we can see that the Zhang-Rice singlet band and charge-transfer characteristics are essentially the same as what we observed in the Fig.4 in the main text  at $U=7, U^{\prime} = 5.6, J_H = 0.7,  n_h=1.045, T=0.08$.

\begin{figure}
  \begin{center}
    \includegraphics[width=0.9\columnwidth]{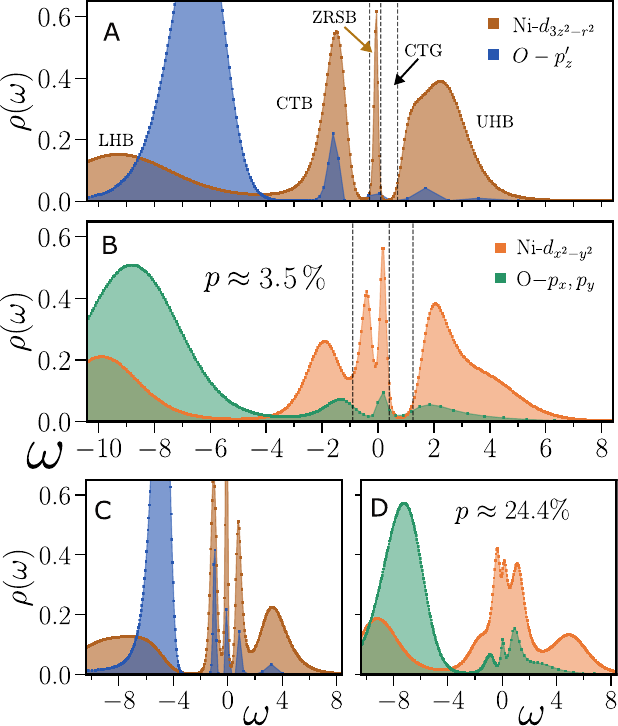}
  \end{center}
  \caption{
 Local density of states of the 11-band Hubbard model of a few  Ni-$d$ and O-$p$ orbitals. Here $U=9, U^{\prime} = 6.3, J_H = 0,  n_h=1.05, T=0.1$. See also Fig.4 of the main text. }
\label{fig:akw}    
\end{figure}

\subsection{Benchmarking the DQMC}

\begin{figure}
  \begin{center}
    \includegraphics[scale=0.65]{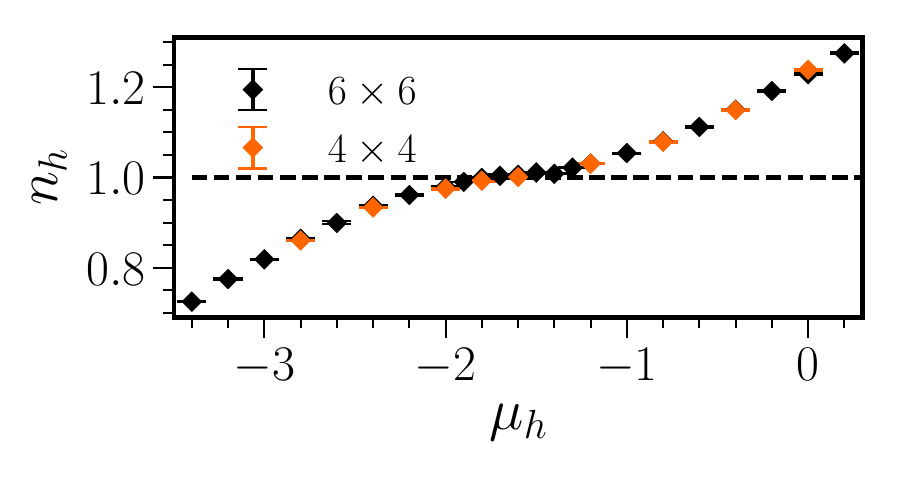}
  \end{center}
  \caption{
{Hole concentration as a function of hole chemical potential. Results are from two different lattice with periodic boundary conditions: $6 \times 6 \times 11$- orbital and  $4 \times 4\times 11$- orbital square lattices. Here $U=7, U^{\prime}=5.6, J_H=0.7, T=0.3$.}
    }
  \label{fig:size}    
\end{figure}

\begin{figure}
  \begin{center}
    \includegraphics[scale=0.7]{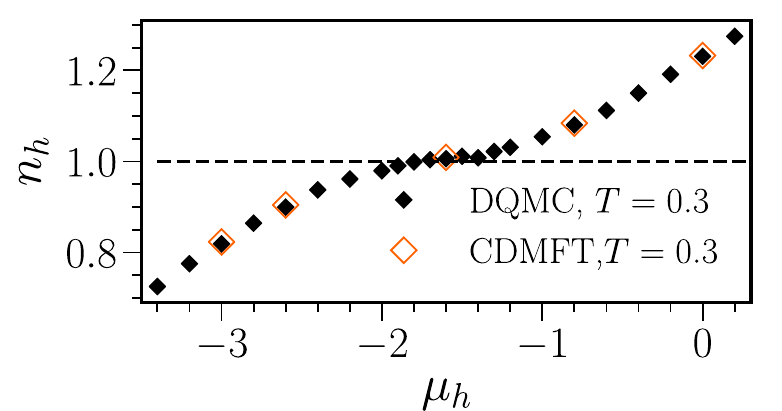}
  \end{center}
  \caption{
   Comparing the CDMFT and DQMC result on the Hole concentration as a function of hole chemical potential. Here $U=7, U^{\prime}=5.6, J_H=0.7, T=0.3$.
   }   
  \label{fig:cmp}    
\end{figure}

In Fig.~\ref{fig:size} we see that for the DQMC simulation at $T=0.3$, changing system size from $6 \times 6 \times 11$ orbitals to $4 \times 4\times 11$  orbitals does not essentially modify our result. Hence,  the finite size effects in our study at these parameters are negligible.  In Fig.~\ref{fig:cmp}, comparison between the CDMFT 
and DQMC result on $n_h$ are shown at temperature $T=0.3$, where good agreement is seen in a wide chemical potential range.

\subsection{Value of superexchange coupling $J_1$ between vertical $\dz$ orbitals}
By diagonalizing the 5-orbital O-$\pz$ - Ni-$\dz$ sub-system  as shown in Fig.~\ref{fig:super}D, we find that the spin-singlet ground state has an energy gain $ \tilde{J}_1 \sim -0.18 eV$ over the spin-triplet states at $U=7eV$. This value can be taken as a crude estimation of the  Heisenberg exchange coupling  $J_1$ between the vertical $\dz$ electrons.   The value of $\tilde{J}_1$ changes with $U$, as shown in Table.~\ref{tab:2}

\begin{table}[h]
\begin{tabular}{ccccc}
\hline \hline 

    $\hspace{0.2cm} U $ & $\hspace{0.8cm}$ $E_0 \hspace{0.9cm} $ & $E_1$ & $\tilde{J}_1 = E_0 -E_1 $\tabularnewline\hline

6 & -35.508 & -35.293 & -0.215
\tabularnewline \hline \hline

7 & -36.697 &-36.519 & -0.178 \tabularnewline \hline 

8 & -37.901 &-37.751 & -0.150 \tabularnewline \hline 

9 & 39.116 &-38.988 & -0.128 \tabularnewline \hline 

\end{tabular}
\caption{\label{tab:hop1} Ground state energy $E_0$ and first excited state energy $E_1$ of the 5-orbital Hubbard model. Here $\mu=0$, $E_{DC} = 2n^{0}_{d_{z^2} }-0.5 \approx 0.65U $. } 
\label{tab:2}
\end{table}







\end{document}